*Chapter 1*

# A GRAPH-BASED PERSPECTIVE TO TOTAL CARBON FOOTPRINT ASSESSMENT OF NON-MARGINAL TECHNOLOGY-DRIVEN PROJECTS — USE CASE OF OTT/IPTV

*Reza Farrahi Moghaddam,*[*] *Fereydoun Farrahi Moghaddam, and Mohamed Cheriet*
Synchromedia Lab, École de technologie supérieure (ETS),
University of Quebec (UduQ), Montreal, QC, Canada

**Keywords:** Sustainability, Green ICT, Footprint, Water Consumption, Energy Consumption, GHG Emissions, Video on Demand (VoD), IPTV, OTT.

**Abstract**

Life Cycle Assessment (LCA) of green and sustainable projects has been found to be a necessary analysis in order to include all upstream, downstream, and indirect impacts. Because of the complexity of interactions, the differential impacts with respect to a baseline, i.e., a business-as-usual (BAU) scenario, are commonly considered to relatively compare various projects. However, as the degree of penetration of a project in the baseline increases, the popular marginal assumption does no longer hold, and the differential impacts may become inconsistent. Although various mythologies have been successfully proposed and used to contain such a side effect, the bottom-up nature, which initiates the assessment from the project itself and ultimately widens the scope, could easily fail to acknowledge critical modifications to the baseline. This is highly relevant in terms of ICT's disruptive and dynamic technologies which push the baseline to become a marginal legacy. In this work, an analytic formalism is presented to provide a means of comparison of such technologies and projects. The core idea behind the proposed methodology is a magnitude-insensitive graph-based distance function to differentially compare a project with a baseline. The applicability of the proposed methodology is then evaluated in a use case of OTT/IPTV online media distribution services.

---

[*]Corresponding author's e-mail address: imriss@ieee.org (Reza Farrahi Moghaddam)

## 1. Glossary of Notations

In this section, before going through the introduction, some key terms and concepts are presented and defined in order to avoid interruption of discussions in the other sections.

- Over-the-top (OTT): A system for the broadband Internet delivery of video and audio without a pay-TV operator being involved in control or distribution of the content.

- IPTV: A proprietary model in which a TV-like service is provided to subscribers on a dedicated access network.

- Subscription video on demand (SVoD): A model that offers subscribers unlimited access to content for a regular monthly, quarterly, or annual fee. It may be referred to as 'narrowcast' compared to the broadcast concept. However, it is worth adding that the narrowcast term can be also used in the context of broadcasting in the form of a 'narrow' transport media multiplexed using switched digital video technology.

- OTT SVoD: A SVoD model that operates over the broadband Internet (for example, Netflix).

- IPTV Broadcast: An IPTV model that delivers TV broadcasts.

- IPTV (S)VoD: A VoD model placed on top of a subscription IPTV model.

- P2P (IP)TV: A peer-to-peer (IP)TV model, which can be seen as a crossover to OTT and IPTV models, and leverages on the 'locality' of the location of the viewers and also 'broadband' penetration in the customer premises in order to emulate an 'IPTV'-like experience without the presence of a dedicated IPTV network [32, 36].

- Binge watching: A form of video watching i) that includes more than one episode of a series or collection in each watching session and ii) through which the watching time is determined by the viewer [20, 27].

- Broadcast: This term could be used in two different contexts. The first one is in the media context, which is actually the focus of this work. The second context could be defined from the standpoint of network. In this perspective, the media provider 'broadcasts' [encrypted] media segments openly to the Internet, and broadcast providers cache such segments depending on their interest. Then, the broadcast providers use the 'cached' content to serve requests from users instead of emulating streaming over HTTP and TCP/IP protocols.

- Time-Shifted TV: Unlike the common live TV model, the programs broadcasted on a TV channel would be available to the viewer at a later time in the time-shifted TV models.

- Graph: A graph $G$ is defined as tuple $G = (V, E, \Omega, U, A)$, where $V$ is a set of its associated vertices (nodes), $E$ is a set of its directed edges, $\Omega$ is a set of weights assigned to edges, $U$ is a set of uncertainties associated with the edges, and $A = A_V \cup A_E$ is a set of attributes associated to either a node or an edge. A particular vertex, edge, weight, uncertainty, and attribute is denoted as $v_i \in V$, $e_{v_i,v_j} = e_{ij} \in E$, $w_{ij}^\alpha \in \Omega$, $u_{ij}^\alpha \in U$, and $\alpha_k \in A$ respectively. Here, $k \in \mathcal{C}_{(V)} \sqcup \mathcal{C}_{(E)}$, where $\mathcal{C}_{(\cdot)}$ is a subset of natural numbers from 1 to the cardinal number of its argument, $\sqcup$ is the disjoint union operator, and $\alpha \in A$. For the purpose of simplicity, we assume a single attribute per object in this work (in addition to the label, in case it is presented). Therefore, in terms of vertices that usually carry a label, the attributes would be the set of their label and possibly another attribute (for example, the geographical region). Besides, in our footprint-related examples, we have $A_E = \{\text{WC}, \text{EC}, \text{GE}\}$, where WC, EC, and GE represent water consumption, energy consumption, and GHG emissions, respectively. It is worth noting that although the weight on the edges could be also encoded as an attribute, it is preferred to have a separate variable for it because of the continuous nature of its values. The weights could be interpreted as 'flow' rates between nodes. The 95% confidence interval of a weight $w_{ij}^\alpha$ is determined by its associated uncertainty $u_{ij}^\alpha$: $\left[w_{ij}^\alpha - u_{ij}^\alpha, w_{ij}^\alpha + u_{ij}^\alpha\right]$. Also, we assume that two graphs share the same level of uncertainty on their common edges.

## 2. Introduction

Sustainability of the world has been a major concern for all stakeholders considering the growing consumption/activity trends comparable with global levels of available resources [12]. The impact of digital activities, usually referred to as ICT, has been suggested to be both positive and negative. Therefore, there have been a considerable number of studies assessing and then leveraging on the advantages while containing negative impacts or footprint [12]. Although the ICT sector can be broken down into three parts, i.e., devices, data centers, and network, there are ICT operations that involve all three parts in an inseparable way. A clear example of these operations is online content services (OnCSs), such as YouTube and Netflix, that require streaming high bandwidth media over wide areas to a large number of asynchronized users.

A diverse number of studies have examined the impact of OnCSs compared with offline approaches (for example, DVD purchase or rental) [25,26,35,39]. These Life Cycle Assessment (LCA) studies have correctly pinpointed many aspects of online media as a 'project' with respect to offline media as the baseline. Many positive enabling effects have been observed and evaluated including a) lower upstream-related (manufacturing) resource depletion, b) less downstream electronic waste (end of life), and c) less transport-related fuel consumption (to/from rental places), among others. Easier access to media can be as well pointed out as a positive social impact. Meanwhile, some negative impacts or concerns have been observed: i) higher electricity consumption associated with playback devices, and also set-top boxes (STBs), ii) higher electricity consumption (EC) associated with network routers and media data centers, iii) higher EC at prime time hours which usually overlap peak electricity consumption hours, and therefore iv) higher ecological footprint (EF), for example GHG emissions, since the grid mix is usually less green at peak hours. Two more concerns to mention would be v) additional EC and EF associated with the media encoding/decoding and also encrypting/decrypting services, and vi) EC and EF associated with playback with no audience.

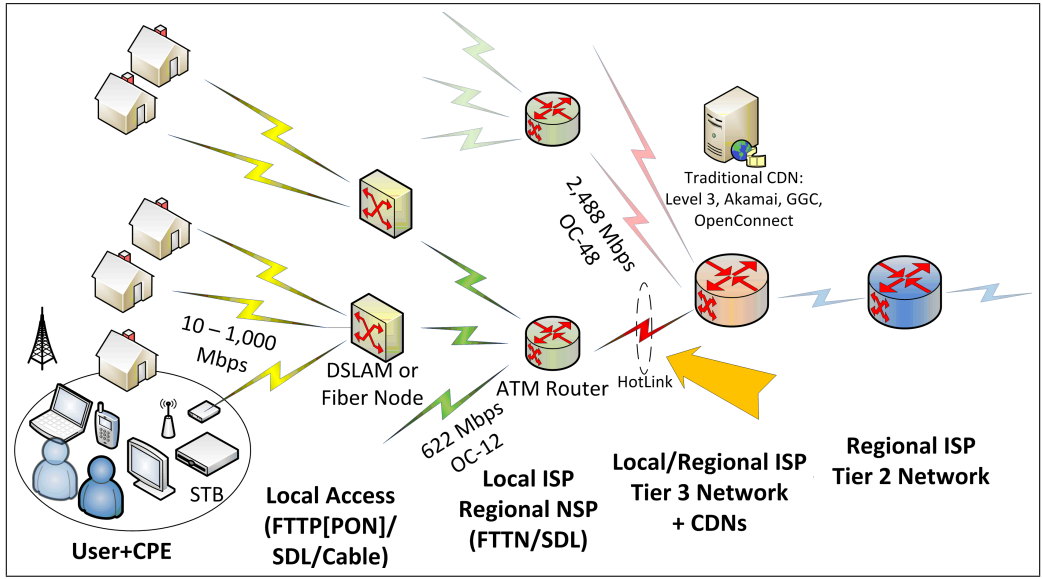

Figure 1. The illustrative diagram of the OnCS case study considered in this work. We assume a penetration of the CDNs down to Tier 3 providers. Therefore, the baseline and projects are defined from that point to the CPE. It will be argued that the high bandwidth OC-48 optical connection would act as a troubling hot link in operation of the OnCS in prime time.

A question beyond the differential impact of OnCSs compared to offline services is what would be the 'impact' of such an impact assessment. This question, which can be informally formulated in the form of 'additionality' requirement, has its roots in the following statement: assuming that the decision to shift to online businesses has been already started and motivated by other factors, such as economic, social, or personal motivations, the added value of the assessment could not be quantified. In such a case, which seems to be the case of OnCSs, the assessment should be focused on the differential impact of two or more ways of the online services with respect to each other. This point is one of the motivations of the current work, and we will discuss it in more detail in section 5., where we define the 'project' and the baseline scenarios.

Another aspect of the OnCSs is complexity of the involved cyber-physical systems and their associated models. Even with the most comprehensive models used, it is necessary to set a 'boundary' beyond which a fixed-state reservoir is assumed. Usually, the spatial and geographical boundaries are easier to identify. However, the most critical boundaries, and especially their associated 'crossing' events [8], would be placed along 'horizontal' temporal and 'vertical' structural dimensions. In particular, when an operation scales up to the scales of society and economy, the 'trend'-based models and the associated boundaries of the external infrastructure entities could easily be violated and 'crossed', and therefore those infrastructure entities are required to be explicitly added to the main model. Although consideration of more comprehensive models, which push the boundary of the system in associated infrastructure and other entities, seems to be a straightforward solution, considerable additional computing costs would force the modeler to stop at some boundaries. It serves as the second motivation of this work to develop a differential assessment metric that does not require explicit models of the 'neighboring' external entities while preserving and enhancing absolutely-small but relatively-strong impacts. To achieve this goal, we propose a graph-based approach to the interpretation of the [differential] assessment of sub-components of a model (please see section 7.). This approach is required because the traditional LCA aggregation would fail to notice those relatively-important impacts that go below the threshold or uncertainly level, or become insignificant compared to bigger impacts at the end of aggregation. We choose to build our approach on top of a graph representation in order to have a high level of flexibility, adaptability, and visualization considering the complexity involved with ICT.

With respect to our case study of OTT SVoD, there is a huge concern regarding OnCSs; the online viewing seems to increase the ratio of 'watching hours' per 'capita/day'. Although the footprint of 'one hour' online watching could be 'less' than that of offline watching, the total footprint per capita, and also per society, would be higher because of more watching hours. A counterargument against this concern would be: the 'spare time' that has been 'managed' and directed by online watching 'activity' could have otherwise resulted in a negative impact since the person would perform 'other' extraversional or introversional activities to fill their spare time which as a result could end up with additional associated footprint at incomparable levels. For example, the depression-related drug costs were estimated to account for 16.65% of the Canada's health expenditures, i.e., B$14.00 out of B$84.10, in 1998 [19]. These costs has been estimated to have been increased to B$35.17 in 2013 [5, 33]. Moreover, the cost of repression has been estimated to be €500 − €2,000 per capita/year in Europe [21, 31]. However, modeling this social impact in the form of behavior modification would require more complex models to account for three major aspects: human, tangible components, and intangible components [17, 24, 34, 36, 39]. Nonetheless, a higher bound of 'potential' footprint adjustment could be associated with well-managed and highly user-engaged OnCSs.

Moreover, the OnCSs could not be independently analyzed from other forms of online interactions, such as 'social networks'. For example, it has been observed that tweets about TV series on Twitter have actually nudged people to watch more 'live' TV compared to time-shifted TV essentially to avoid spoilage and also to keep up with online and real-time commenting activities related to the content [38]. However, it can be argued that this trend mostly works for 'a few' series that are highly popular in the society, and the trend for the rest of the programs would be quite the contrary, i.e., more time-shifted, video on-demand, and binge watching compared to live watching.

Finally, it is worth mentioning that time-shifted TV does not always mean less EC and EF. For example, hardcoded time-shifted channels on satellite TV would have 'multiple' counts of transmission footprint because of transmitting the 'same' signal over several channels rather than transmitting one channel and locally timeshifting the content in the DVR/PVR at the customer premises. We will address such cases in the future.

The specific case study of this paper is shown in Figure 1. We argue that a point of crisis in the OnCSs, especially for those which have used CDNs,[1] is the connection from the Tier 3 NSPs to the local ISPs. These high-bandwidth connections, such as OC-48 with a capacity of 2,488.32 Mbps, usually serve a large number [thousands] of end users. In our case study, we consider a regional span of 5,000 end userrs served by an OC-48 connection as the baseline. The specs are provided in Table 1.

| Specs | Optical Link | Capacity (Mbps) | Max Number of 5-Mbps streams | Avg Mbps for 5,000 streams | Monthly Lease Cost |
| --- | --- | --- | --- | --- | --- |
| Baseline | OC-48 | 2,488.32 | 497 | 0.5 Mbps | $200,000 |

Table 1. The specification of a local ISP's backhaul optical connection. This connection is the starting point for the baseline and also the projects in our case study.

The paper is organized as follows: the complexity of the influential factors associated with OnCS is discussed in section 3. Then, In section 4., the assessments of various components of a typical OnCS is presented. The project(s) and baseline of this study are finalized in section 5.; two projects are considered that leverage on the micro-registration concept of content streams to reduce EC and EF compared to a baseline of unicast OnCS. It is followed by a discussion on similarities and differences between the problems at hand and that of on-line communications, such as IP Multimedia Subsystem (IMS). In section 7., the proposed graph-based distance metric of the assessment is presented. The application of the proposed approach to the case of OnCS projects is provided in the next section demonstrating the fact that despite no-impact nature of the micro-registration technique in many of the big components of the OnCS, the proposed approach was able to successfully spot the potential capital footprint associated with ISP's backhaul congestion. Finally, the conclusions and some future prospects are provided in section 9.

## 3. Influential Factors Involved in an OnCS

The complexity and variability of the OnCSs could undermine any conclusion and assessment if influential factors are not considered. In this section, some of the key factors involved in many OnCSs are listed. However, since we could not cover all of them in this work, they will be considered in more depth in the future.

### 3.1. The 'unit' of provided service

The question of proper unit of service or product is a critical and common question in all assessments, especially in case of life-cycle assessment (LCA). However, the unit of service

---

[1] Such as Google's Google Global Cache, Netflix's OpenConnect, Level 3, and Akamai.

could impose a considerable level of complexity in the case of OnCSs.

As mentioned in the introduction, the ratio of 'watched hours' per 'capita/day' (WH/CD) could be a reasonable unit of service that can implicitly absorb a considerable number of social-related effects. However, it can be argued that this unit does not consider the EC and EF associated with streaming the 'blackholed' content to the customer premises (delivered content with no audience).[2] This is particularly important in the case of unicast versus multicast services. For example, it has been shown that multicast IPTV broadcast models have less EC and EF compared to their unicast counterparts [6, 24, 29, 30]. In case of the present study, which is OTT SVoD, the streams are initiated per request and are ephemeral. Therefore, we argue that the ratio of watching hours per capita/day is an appropriate unit of service in our case despite the fact that we consider unicast baseline versus multicast projects.

As a side note, we would like to raise a concern regarding the level of inequality in societies [9, 11]. Although for societies with high score of Inequality-adjusted Human Development Index (IHDI) [1], averaged values are well representing the status of the society; other approaches should be considered when a low value of IHDI is involved. In other words, excessive level of service to a small portion of population would mask out the low level of service provided to the rest of the society in a simple averaging approach. Such a society is not eligible to receive the associated credits of that service level. In future work, using a 'threshold-then-average' approach we address this problem. In this study, the involved society is highly localized with a high score of IDHI, and therefore average values are acceptable.

### 3.2. To increase or not to increase

Although this point requires additional studies, it could be argued that there is no immediate conclusion on whether to increase or decrease the watched hours per capita/day. In particular, the impacts on mental health, education, integration of society, and EF reduction should be simultaneously considered.

### 3.3. Overpriced online content

Assessment of an OnCS would require special attention because of the potential multi-business nature of the associated firm. Some of these businesses, which could be offline content service providers of the same content provided on the OnCS, would push the firm to put 'overpriced' tags on the content when delivered by the OnCS. The overpricing business would go unnoticed because the users have been used to the same price when accessing the offline services. Nonetheless, the price of the offline services would considerably include the physical media used to carry the content (such as DVDs). Therefore, overpricing can be practically considered as the un-neutrality of the firm in handling its multiple businesses, while promoting its online business as a sustainable 'fashion'. This way of business, which would have its root in the revenue model of the firm, would impact and slow down the adaptation of dematerialized services and adaptation of ICT in general. Therefore, the firm would not be eligible to collect the whole sustainability credits of its OnCS business. One way to implement this requirement would be to assess the footprint of a 'content' (for example, a movie) instead of its 'content forms' (such as that movie on DVD, online, among others).

---

[2] 'Interfaces', especially human-computer interfaces (HCI), could present a considerable challenge and also potential to a sustainable approach to ICT [12].

### 3.4. FTTH, smart grid, broadband Internet access

Another aspect related to streamed OTT SVoD operations is the broadband access to the customer premises, which spans along a path handled by many actors from content provider, CDNs, transit networks, retail ISP, and even telephony firms. Although the questions of peering and congestion are popular in this regard [8, 14, 22], another aspect related to fiber deployment and associated footprint is also of interest in this section.

Usually the connections, especially the last-mile cable or fiber to the customer premises, is deployed and amortized by the ISPs. However, with the recent move toward smart everything, from smart grid, smart house, to smart city, there has been a need for broadband connections to the customer house, such as fiber-to-the-house (FTTH) or fiber-to-the-premises (FTTP) deployments, in order to host the bandwidth-hungry telemetry links from the sensors and appliances at the customer's house to the smart management centers and clouds.[3] The extra bandwidth available on such subsidized connections has prompted delivering OTT broadband services on the same fiber. With focusing only on the EF, it seems a necessity to allocate some of the 'capital' EF associated with the FTTH deployments to the ONCS' EF. In other words, the argument, that the share in the capital EF of a broadband Internet connection would be negligible for an individual service because of the large number of services that would use that connection, would not hold because the excessive bandwidth would not be fully utilized except for the OnCS and other similar services. To be more precise, it could be the profitability of the OnCSs that drives wide deployment of FTTH.

In the present case study, the impact of the capital EF of connections, especially at the regional ISPs' bottleneck, is of interest. It will be shown that even without an explicit model, the proposed graph-based distance could spot and highlight the risk of requiring to deploy extra fiber links, exchange points, and routers to handle exponentially growth in bandwidth required for the content streams.

### 3.5. The baseline is not baseline anymore

It is worth noting that the nature of ICT and its associated advances in terms of technology has resulted in a high level of ambiguity in the definition of the 'baseline'. In particular, there is a great potential for efficiency, and reducing EC and EF that might have been neglected in the past because of their non-profitability. In other words, the baselines and BAUs usually referred to are those concerning profit not those in terms of 'physics' of the systems involved. In addition, many technologies with less EC and EF become practical every year, and would continuously redefine the baseline. It seems as if it is a necessity that the assessments challenge the baseline in order to create 'guidelines' for the future disruptions that would anyway occur because of profit-driven mechanisms, even if the assessments could not move the baseline themselves.

---

[3]Possibly in the form of municipal broadband providers.

### 3.6. Distributed, peer-to-peer, and immersed content

It is worth noting that, in continuation of the discussion in the last subsection, other forms of content 'containers', such as i) distributed, ii) peer-to-peer, and iii) immersed data storage [8], could provide big shifts in the baseline and therefore in the EC and EF of OnCS businesses. In particular, the network's EC and EF associated with streaming of the content, in both form of capital and operational, could be drastically cut down.

### 3.7. Differential versus Marginal Assessment

As mentioned in the introduction, the focus of this work is on development of a distance between a project and its baseline. The proposed graph-based distance, which will be presented in section 7., could be also used to calculate the differential distance between two projects with respect to a common baseline. In either case, we argue that a marginal analysis would be misleading, and a differential one is necessary as will be provided by the proposed distance.[4] In this subsection, the necessity of a differential assessment against a marginal one is discussed.

It has been argued in some studies that the selection of marginal over 'non-marginal' average EF is not always the best choice, and it has been suggested to at least report both values [18, 28]. In particular, the first users have been exposed to smaller levels of resource scarcity than the marginal ones, and therefore a marginal analysis would not be fair to the newcomers. This is especially important in the case of disruptive technologies and businesses even if the depleted resource has an upper bound capacity (for example, water resource). Any disruption to the 'baseline' way of living and business in the societies has been usually considered as a negative change. However, many benefits of sustainable actions, such as those of dematerialization, ICT takeover, and especially virtualization, would not be 'positive' if they are not implemented at disruptive scales. A shallow integration of ICT and artificial intelligence (AI) in the society has resulted in avoidance of many realistic challenges that ICT would face given the fact it would be planned to take over a considerable portion of materialized activities.

Even in the case of non-disruptive businesses, it seems that the marginal analysis would be unfair because it would prevent expansion of 'green' resources, such as green electricity, in the regions that have 'dirty' marginal sources. Therefore, in a long term vision, it seems that marginal analysis would be less practical in either case of disruptive or non-disruptive changes. In contrast, a differential analysis would be interesting in spotting the advantages of a project, and a fine-grained differential distance, as will be introduced in section 7., it could provide insights on small changes that would make big differences when their associated external entities would change their state in a nonlinear way. A differential analysis would not prioritize first users against the newcomers who would be mostly ICT businesses.

---

[4]It is worth nothing that differential analysis has been used before, for example in [37].

## 4. Data Sources & Basic Footprint of Components

As mentioned in the introduction, we propose a new interpretation approach to footprint assessment in this work. This approach starts from the evaluated footprint assessment of sub-components of the baseline and also those of the project, and then develops a graph-based differential 'picture' that finally is translated in the footprint difference between the project and the baseline. This approach can be directly applied to two projects in order to generate their differential footprint with respect to each other even in the absence of an explicit baseline. In this section, we use the state of the art in order to collect the footprint assessment of the sub-components of an OnCS to be fed in the proposed approach in subsequent sections. Various work have studied online media delivery services and their footprint [25, 26, 35, 39]. In order to have consistency within the footprint values, we only use those reported in [25] from here on. Regression and projection of all these studies will be considered in the future.

In the previous section, it was argued that the best unit of function or operation for the OnCSs would be 1 Watched Hour per Capita per day (1 WH/CD). This unit simply excludes the social and other influential factors of the OnCSs that would promote new watching behaviors such as binge watching, and therefore allows a comparison at the delivery system level. Considering the scenario that was presented in the introduction on prime time watching of a popular series in the form of OTT SVoD in an area covered by an OC-48 regional connection, each video stream would require a download volume of 2.2 GB for 1 WH/CD (with a bandwidth of 5 Mbps). Using the data provided in [25], this download volume would have the associated EC and GE footprint that are presented in Table 2. It is worth noting that the grid mix used is that of Sweden yearly averaged. However, it can be mentioned that the grid mix at 6PM, i.e., the prime time hour, would have much associated footprint, and an extra factor should be considered [4, 7]. We will consider this fact in the future. Also, for the User and CPE devices, it is assumed that a lifetime footprint of 600 kgCO$_2$e of the devices is amortized over a working lifetime of 9,000 hours one third of which is allocated to the OnCS activities. The associated EC embodied in these devices is calculated using the averaged Sweden grid mix's emission factor.

| Component | EC (KWh per WH/CD) | GE (kgCO$_2$e per WH/CD) |
|---|---|---|
| Non-local access network (use phase) | 0.176 | 0.011 |
| Non-local access network (embodied) | 0.587 | 0.035 |
| Local access network (use phase) | 0.176 | 0.011 |
| Local access network (embodied) | 0.587 | 0.035 |
| User and CPE (use phase) | 2.860 | 0.172 |
| User and CPE (embodied) | 0.370 | 0.022 |

Table 2. A summary of the EC and GE associated with various parts of the baseline scenario. GE is calculated using the *averaged* emission factor of the Swedish electricity grid, i.e., 0.06 kgCO$_2$e/kWh.

## 5. The 'project'

As mentioned in section 3.1., we consider the ratio of watched hours per capita/day as the unit of service in our case study. However, it is required to identify and set the baseline and project. It seems that setting offline content services, such as DVD rental, as the baseline against an OnCS as the project could be a straightforward choice. However, with the disruptive shift of business to OnCS regardless of the associated EC and EF, which would raise some additionality-related concerns, it seems that it is more practical and fair to define the baseline and project from the set of OnCSs. The comparison with DVD would not bring much benefits because the disruptive transition has been already started.

In other words, it does not matter if the online way of business has 'less' footprint compared to the offline counterpart. Instead, what is important is how to minimize its footprint to "zero" while maximizing its benefits to "infinity". Being 'better' is no longer a green light for a society-scale project, and only the 'best' sustainability-oriented projects should be 'designed' and pursued. However, because of limited 'knowledge' and also practical 'constraints', an actor or a collection of them would promote and implement an intermediate project in the short term. It is important that the outcome of these projects are collected in the form of 'lessons learned' and 'bottlenecks identified'. Many constraints, especially those that root in the 'legacy' effect of giant low-level software, would not be addressed or softened if the 'sustainable' project do not acknowledge their own weak points. As embarrassed by many actors, the next step could be the 'open' movement: open data, open knowledge, and open 'software' in order to lift constraints in a crowd-sourced manner.

For this work, we consider the baseline to be unicast OTT SVoD. Therefore, in the baseline, each connected viewer receives an individualized stream all the time from the 'server' of the provider. In practice, the server would be that of the provider's CDN closest to the viewer. We consider two 'projects' against this baseline:

### 5.1. Project 1: Micro-Registration (Micro-Reg or MR)

In our case study of a prime time OTT SVoD scenario, we assume that the end users asynchronously start their viewing in an interval of 30 minutes starting from 6PM. In other words, it is assumed that there would be 5,000 independent requests submitted within that period. Regardless of the quality of the delivered video, the full bandwidth of 2,488.32 Mbps will be utilized in the baseline scenario.

As the first project to improve the performance and also to reduce the footprint, we propose to implement a 'micro-registration' strategy in order to reduce the number of concurrent streams delivered to the end users. In the micro-registration strategy, an artificial delay is added to received requests in order to align them together in bunches of multicast streams placed in intervals of 5 seconds with respect to each other. It will be shown that the continuous superior quality of the delivered content would be more than 10 times higher than that of the baseline, and therefore it could be argued that a few second delay at the start of the viewing session would be bearable by the end users. Considering the specs of our case study, we would require only 360 unique concurrent streams to serve all users, and therefore an actual bandwidth of 6.9 Mbps per stream could be achieved which is even higher the recommended bandwidth for HD programs, i.e., 5 Mbps. It is worth mentioning that this is achieved without any change to the hardware and installations, and it only requires smart management of requests and also ability to multicast in local access part of network. The footprint of 1 WH/CD associated with the Micro-registration Project (MRP) is provided in Table 3. As can be seen, the only difference is a minor use-phase footprint reduction at the non-local access network. We did not apply the reduction to the embodied footprint because the hardware at the prime time is implicitly allocated to SVoD service whether it is used or not. In section 8., it will be shown that this small footprint can be enhanced and spotted by the proposed graph-based distance in order to avoid footprint associated with implementation of extra links. Such avoidance is possible with smart management strategics, for example the two proposed projects in this section. The second project is described in the following subsection.

| Component | EC (KWh per WH/CD) | GE (kgCO$_2$e per WH/CD) |
|---|---|---|
| Non-local access network (use phase) | **0.128** | **0.008** |
| Non-local access network (embodied) | 0.587 | 0.035 |
| Local access network (use phase) | 0.176 | 0.011 |
| Local access network (embodied) | 0.587 | 0.035 |
| User and CPE (use phase) | 2.860 | 0.172 |
| User and CPE (embodied) | 0.370 | 0.022 |

Table 3. Compared with Table 2, this is the summary of EC and GE associated with the micro-registration project. The modified values are highlighted.

### 5.2. Project 2: Delayed Micro-Registration (Del-Micro-Reg or DMR)

The aim of the Delayed Micro-registration Project (DMRP) is to increase the time interval between consequent streams up to 15 seconds, which would allow a three times gain compared to the original Micro-registration Project. Such a delay is achieved in a multi-step process described below. With every new request received, the request is instantly served by the 'live' broadcast of the program. In this way, there is no delay in viewing, however, there is a time shift with respect to the beginning of the program. This step gives an experience similar to that of 'live' TV to the viewer. There is a chance that some viewers would continue watching the program on the live stream skipping the missed part. For the rest of the new users, they are linked to the latest stream. Considering the 15-second intervals between the streams, the viewer would miss a few seconds of the beginning of the program up to 15 seconds. If there are some users that are not still satisfied, they are linked to the next stream, and they could watch the program from its beginning. The three steps of the proposed project would make it bearable to wait for a long interval of 15 seconds in rare cases to start viewing a program.

Similar to the Micro-registration Project, the EC and GE of the Delayed Micro-registration Project are provided in Table 4. Again, there is only some reduction in the use-phase footprint and consumption associated with the non-local access network. The comparison between the baseline and the projects in terms of quality of service, EC, and GE is provided in section 8.

| Component | EC (KWh per WH/CD) | GE (kgCO$_2$e per WH/CD) |
|---|---|---|
| Non-local access network (use phase) | **0.043** | **0.003** |
| Non-local access network (embodied) | 0.587 | 0.035 |
| Local access network (use phase) | 0.176 | 0.011 |
| Local access network (embodied) | 0.587 | 0.035 |
| User and CPE (use phase) | 2.860 | 0.172 |
| User and CPE (embodied) | 0.370 | 0.022 |

Table 4. Compared with Table 2, this is the summary of EC and GE associated with the delayed micro-registration project. The modified values are highlighted.

Note that our case is different from that of [29, 30] because in our case we consider VoD that could have a high degree of time shift introduced by asynchronized behavior of viewer to connect and request a program. This is in contrast to the case of [29, 30] that considers TV broadcasting and therefore a synchronized program is always provided to the viewer from a 'pool' of channels.

## 6. A note on similarity with the case of OTT Video Calling

Although the nature of content exchanged in the SVoD is highly different from that of video calls, there is a chance to generalize the concept of micro-registration to personal content delivery services thanks to the fact that personal content exchanges are shifting from 1-to-1 models to 1-to-N models instead. This behavioral change, which highly correlated with the integration of social networks in personal lives, provides a great potential to apply multicast strategies to personal content exchange scenarios. However, this would require more modifications in terms of management and also seamless integration of social networks and Telco operators.

## 7. Proposed Graph-Difference Approach to Impact Assessment

In this section, the proposed approach to calculate the distance between two scenarios (for example, a baseline and a sustainable project) is provided. The proposed approach is based on a generalized graph distance which considers the weight on the edges in the calculations of the distance. In addition, we introduce an attribute-sensitive graph reduction method that allows preserving those otherwise similar vertices that carry different attributes. It will be shown that the well-known LCA approach to the footprint analysis and assessment can be modeled in the language of the proposed graph-based approach.

Comparison of graphs has been well studied in various fields of application [3, 16]. A common approach to distance between graphs has been based on the 'size' of graphs and their union and intersection [13, 40]. In particular, a graph distance $d(G_1, G_2)$ has been defined [40]:

$$d(G_1, G_2) = 1 - \frac{m(G_1, G_2)}{M(G_1, G_2)}, \tag{1}$$

where $m(\cdot,\cdot)$ and $M(\cdot,\cdot)$ are measures of similarity between two graphs and size of the problem, respectively:

$$m(G_1,G_2) = \|G_{12}\|, M(G_1,G_2) = \|G_1\| + \|G_2\| - \|G_{12}\|, \tag{2}$$

where $G_{12}$ is the maximum common subgraph of $G_1$ and $G_2$, and $\|G\|$ is the size of the graph $G$ that is defined as the number of its vertices. Below, we define a generalization of the graph distance function that considers topology, weights, and attributes.

### 7.1. Weighted Graph Distance

Here, the proposed weighted graph distance is defined based on a generalized difference graph.

**Definition 1** (Weighted Difference Graph). The difference graph $\delta(G_1,G_2)$ between two graphs $G_1 = (V_1,E_1,\Omega_1,U_1,A_1)$ and $G_2 = (V_2,E_2,\Omega_2,U_2,A_2)$ is defined by:

1. **Initializating $\delta(G_1,G_2)$:** Define the initial $\delta(G_1,G_2)$ to be the subgraph of $G_1$ that consists of a differential edge set $E_1 \setminus E_2$ and its associated vertices.

2. **Substrating the edges:** For each edge $e_{ij}$ in the initial $\delta(G_1,G_2)$, set the weight equal to $w^\alpha_{1,ij} - w^\alpha_{2,ij}$, where $w^\alpha_{1,ij}$ is the associated weight of the edge $e^\alpha_{ij}$ on $G_1$, for example.

3. **Pruning zero-weighted edges:** Remove all edges that have a zero or negative weight.

4. **Pruning isolated vertices:** Finally, drop all those vertices that have no edges connected to them.

The final graph following these steps is defined as the weight difference graph $\delta(G_1,G_2)$. Figure 2 shows an example of calculating the difference graph.

Before defining the weighted graph distance, two other measures are defined: i) The weighted graph size and ii) the relative weighted graph size:

**Definition 2** (Weighted Graph Size). The weighted size of a graph $G = (V,E,\Omega,A)$ is defined to be the sum of its weighted vertex-size and its weighted edge-size:

$$\|G\|_\Omega = \|G\|_{\text{vertex},\Omega} + \|G\|_{\text{edge},\Omega}, \tag{3}$$

where

$$\|G\|_{\text{vertex},\Omega} = \sum_{i \in C_V} \sum_{\alpha \in A} w^\alpha_{\text{vertex},i}, \tag{4}$$

$$\|G\|_{\text{edge},\Omega} = \sum_{i,j \in C_V} \sum_{\alpha \in A} \left(w^\alpha_{ij} + u^\alpha_{ij}\right), \tag{5}$$

$$\tag{6}$$

Here, $w_{\text{vertex},i}$, which is defined as the associated vertex weight of $i$, is the sum of all the weights of every 'outbound' edge originated from the vertex $i$ multiplied by the total vertex weight of the associated 'sink' vertex plus one:

$$w^\alpha_{\text{vertex},i} = \sum_l \left(w^\alpha_{l_i,l_j} + u^\alpha_{l_i,l_j}\right)\left(w^\alpha_{\text{vertex},l_j} + 1\right), \tag{7}$$

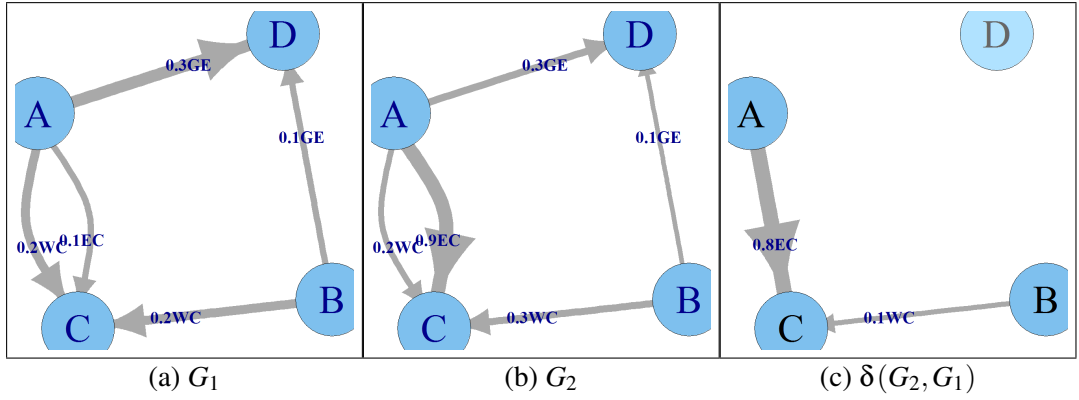

Figure 2. An illustrative example of the proposed graph difference. a) and b) The two given graphs: $G_1$ and $G_2$, and c) The graph difference $\delta(G_2, G_1)$.

where $l$ counts on the outbound edges originated from $i$ having attribute $\alpha$, and $l_i$ and $l_j$ are the source and sink vertices of the edge represented by $l$. It is obvious that those edges, which sit closer to the 'zero-outbound' vertices, have the most influence in determining the size of the graph because their weights would propagate through all those upfront vertices being sourced toward these edges. Although the proposed graph size is general, in our use cases we consider three zero-outbound vertices that correspond to three impact categories of WC, EC, and GE, which actually form the attribute set of edges. It is interesting to mention that $\|G\|_{\text{edge},\boldsymbol{\Omega}}$ is not sensitive to the arrangement of the edges and their weights, while $\|G\|_{\text{vertex},\boldsymbol{\Omega}}$ is highly dependent on how the vertices are connected to each other. It is worth mentioning that the boldface $\boldsymbol{\Omega}$ symbol is used to show the weighted nature of the proposed size, and it should not be mistaken with the actual $\Omega$ set of the graph $G$.

**Definition 3** (Relative Weighted Graph Size). The relative weighted size of a graph $G = (V, E, \Omega, A)$ with respect to a 'reference' graph $G_0 = (V_0, E_0, \Omega_0, A_0)$ is defined in a similar way to that of the weighted size:

$$\|G\|_{\boldsymbol{\Omega}, G_0} = \|G\|_{\text{vertex},\boldsymbol{\Omega}, G_0} + \|G\|_{\text{edge},\boldsymbol{\Omega}, G_0}, \tag{8}$$

where

$$\|G\|_{\text{vertex},\boldsymbol{\Omega}, G_0} = \sum_{i \in C_V} \sum_{\alpha \in A} w^{\alpha}_{\text{vertex}, G_0, i}, \tag{9}$$

$$\|G\|_{\text{edge},\boldsymbol{\Omega}, G_0} = \sum_{i,j \in C_V} \sum_{\alpha \in A} \left( \frac{w^{\alpha}_{ij} + u^{\alpha}_{ij}}{w^{\alpha}_{G_0, ij} + u^{\alpha}_{ij}} \right), \tag{10}$$

$$\tag{11}$$

Here, $w_{\text{vertex}, G_0, i}$, which is defined as the associated relative vertex weight of $i$, is the sum of all the 'relative' weights of every outbound edge originated from the vertex $i$ multiplied by the relative vertex weight of the associated sink vertex plus one:

$$w^{\alpha}_{\text{vertex}, G_0, i} = \sum_{l} \left( \frac{w^{\alpha}_{l_i, l_j} + u^{\alpha}_{l_i, l_j}}{w^{\alpha}_{G_0, l_i, l_j} + u^{\alpha}_{l_i, l_j}} \right) \left( w^{\alpha}_{\text{vertex}, G_0, l_j} + 1 \right), \tag{12}$$

It can be observed that the relative sizes show the impact of a 'significant' change on an edge while the other sizes including the weighted size defined by the equation (3) could not capture such a change. Therefore, we will use the relative weighted size in the definition of the graph distance as shown below. For the purpose of simplicity, from here on, we assume all uncertainty values are zero: $u_{i,j}^\alpha = 0$.

**Definition 4** (Weighted Graph Distance). The proposed graph distance $d(G_1, G_2)$ between two graphs $G_1 = (V_1, E_1, \Omega_1, A_1)$ and $G_2 = (V_2, E_2, \Omega_2, A_0)$ is defined as follows:

$$d_\Omega(G_1, G_2) = \frac{1}{2}\left(\|\delta(G_1,G_2)\|_{\Omega,G_1} + \|\delta(G_2,G_1)\|_{\Omega,G_2}\right), \tag{13}$$

Although the definition of the proposed graph distance is symmetric, we will examine its eligibility to be a metrics in future work.

In the case of our illustrative example, the weighted graph distance can be calculated as follows using the difference graphs already calculated and presented in Figure 2:

$$\|\delta(G_1,G_2)\|_\Omega = 0, \tag{14}$$
$$\|\delta(G_1,G_2)\|_{\Omega,G_1} = 0, \tag{15}$$
$$\|\delta(G_2,G_1)\|_\Omega = 1.8, \tag{16}$$
$$\|\delta(G_2,G_1)\|_{\Omega,G_2} = 1.22, \tag{17}$$
$$d_\Omega(G_1,G_2) = \frac{1}{2}\left(\|\delta(G_1,G_2)\|_{\Omega,G_1} + \|\delta(G_2,G_1)\|_{\Omega,G_2}\right)$$
$$= \frac{1}{2}(0+1.22) = 0.61. \tag{18}$$

## 7.2. Attribute-weight-sensitive Vertex Merge

With the present computing power available [2, 23], graph reduction for medium size problems is not a feasibility requirement to perform an analysis; however, reducing the size of a graph is still of interest mainly because of i) providing 'visual analytics' at the computational power capacities of human, and ii) high-level, automatic interpretation. Various approaches to perform node merging and graph reduction have been studied in the literature [15]. Insensitivity with respect to the attributes of the edges has resulted in aggressive reduction methodologies that have significantly simplified complex problems. However, in our case, the critical points seem to be those that would be easily aggregated and merged within bigger ones. Therefore, we introduce a generalized node (vertex) merge process that respects the 'attributes' of the vertices and edges in the hope of preserving those potential default points in the final 'interpreted' analysis.

**Definition 5** (Node Merge). For a graph $G = (V, E, \Omega, A)$, an attribute-weight-sensitive node merge operation for two vertices $v_i$ and $v_j$ is defined as follows:

**1 Validating the admissibility of the candidate vertices:**

1. Vertex attributes: The attributes of the two vertices (expect their labels) should be the same.

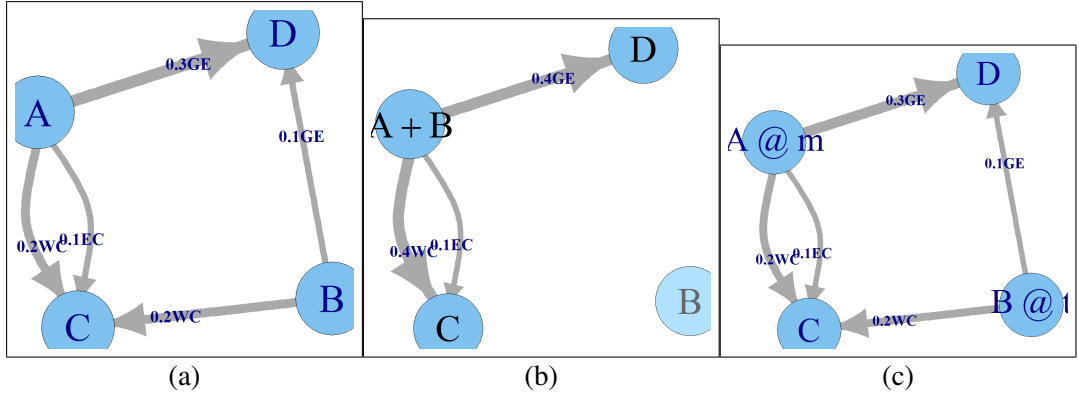

Figure 3. An illustrative example of the proposed graph merger. a) The given graph $G_1$, b) The associated merged graph, and c) The graph cannot be merged when attributes of the vertices do not match.

   2. Edge attributes: The edges of a common attribute should link to the 'same' sink vertex.

**2 Merging the edges:** For each pair of edges of a common attribute, a single edge of the same attribute is created that carries the sum of the weights of the paired members.

**3 Merging the vertices:** A new vertex is created to represent the two vertices. The attributes are already the same. The label is generated by a concatenation operation of the merging vertices' label. The merged edges, created in step (**2**), have their source linked to the new vertex.

**4 Clean up:** All original edges of the merging vertices and the vertices themselves are deleted.

It is worth mentioning that the proposed merge operation can be easily generalized to more than two vertices.

The node merge operation can be performed on as many pairs of compatible vertices as possible. The study of the impact of order in such a process will be considered in future work. Figure 3(b) shows an example of the proposed node merge operation on the graph of Figure 2(a). It is worth mentioning that the merged vertices, i.e., A and B, do not carry any attribute except their labels, and therefore they are admissible for a merger. In contrast, if the vertices A and B carry a different attribute with respect to each other, the merger would default. Such a case is shown in Figure 3(c). The attribute of every vertex is shown after the @ symbol along its label.

## 8. Analysis and Discussion

Before discussing our case study, it is worth mentioning that the graphs in the proposed model should not be mistaken with those that would appear in Multi-Region Input-Output (MRIO) or Multi-Entity Input-Output (MEIO) analyses [10]. More precisely, the graphs addressed in this work are the outcome of an analysis and assessment step which could have been carried out by means of LCA, MRIO, or MEIO analyses (and possibly a combination of them). The proposed graph distance method receives these 'raw' assessments, represents them in the form of a graph, and then provides a novel 'interpretation' that would highlight bottlenecks and critical points which would have been missed if processed through the medium of traditional aggregation.

Extended analyses, such as those of MRIO or consequential LCA, provide insights into probable consequences that would not be captured in the direct footprint; however, they require a 'model' for every component involved. Even then, there is a chance that a disruptive change has been forgotten. Although models are usually nonlinear, we call this phenomenon 'linear side-effect' of modeling. In other words, the linearity is within the methodological approach or simply mentality of the 'modeler' themselves. A straightforward example of the linear side-effect in our case study is the potential unaccounted footprint associated with additional deployment of high bandwidth regional network pipes in the analyses.

The impact of the two proposed projects in terms of quality of delivered video can be observed from Table 5. Although the averaged dividend bandwidth for each viewer is a low value of 0.5 Mbps, the delivered bandwidths are 6.9 and 20.7 Mbps for the proposed projects respectively thanks to micro-registration of service requests. Both these bandwidths are higher than the recommended bandwidth of 5 Mbps, and therefore they would mean a higher quality of service (QoS) and quality of experience (QoE) associated with the 'delivered' online content. Below, we will see the EF of these projects is also less. This contradictory result shows the importance of the 'smart' management in the ICT activities, which can be then interpreted as a fact that the level of state-of-the-art efficiency and optimality in ICT, especially in software components, is still very low.

| Specs | Optical | Capacity (Mbps) | Number of req. unique streams | Avg alloc. Mbps for a viewer |
|---|---|---|---|---|
| Baseline | OC-48 | 2,488.32 | 5,000 | 0.5 Mbps |
| Project: Micro-Reg | OC-48 | 2,488.32 | 360 | 6.9 Mbps |
| Project: Delayed Micro-Reg | OC-48 | 2,488.32 | 120 | **20.7** Mbps |

Table 5. The comparison of the baseline and the projects in terms of the bandwidth and quality of delivered media.

The graph of the baseline and its associated merged graph are shown in Figure 4. We use a '$_{\text{-Mgd}}$' tag to identify the merged graphs. The associated graphs of the two projects are shown in Figures 5(a)-(b). The associated merged graphs are provided in Figures 5(c)-(b). Using these graphs, the difference graphs between the baseline and the projects are also provided in Figures 6(a)-(d). Also, the difference graphs between the two projects are provided in Figures 6(e)-(f). As can be seen, the differences are small in terms of their values. Using the definition of the proposed graph-based distance, i.e., equation (13), the distance among the baseline and the projects can be calculated as provided in Table 6. For the purpose of comparison, the distances are also calculated using the traditional aggregation and the traditional graph distance approaches. As can be seen, the aggregation approach is unable to observe the small but critical differences, and the traditional graph distance only considers the structural differences. In contrast, the proposed distance provides a balance between differences in the weights and also structure. In particular, when this distance is applied to the unmerged graphs, it provides a highly sensitive analysis of the critical but low value differences that can be pursued in order to avoid any large scale change to the external components usually ignored in the original modeling. As an example, the graph-based distances between the baseline and the micro-registration project are explicitly calculated in the Appendix A.

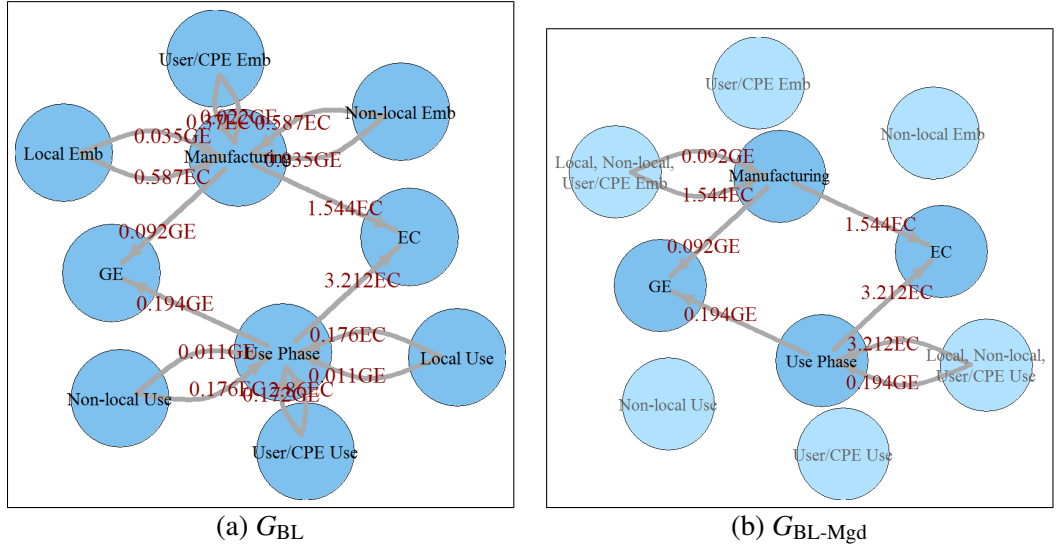

(a) $G_{\text{BL}}$       (b) $G_{\text{BL-Mgd}}$

Figure 4. The footprint graph of the baseline scenario in our case study. a) The footprint graph and b) The associated reduced graph when the graph merge is applied.

| Specs | Traditional Aggregation | | Proposed Graph Representation | | | |
|---|---|---|---|---|---|---|
| | | | Traditional Graph Dist | | Proposed Graph Dist | |
| | Merged | Unmerged | Merged | Unmerged | Merged | Unmerged |
| Baseline vs Micro-Reg | 1.5%EC 1.6%GE | 1.5%EC 1.6%GE | 0.667 | 0.571 | *0.031* | **0.581** |
| Baseline vs Del-Micro-Reg | 4.1%EC 4.1%GE | 4.1%EC 4.1%GE | 0.667 | 0.571 | *0.083* | **1.596** |
| Micro-Reg vs Del-Micro-Reg | 2.7%EC 2.7%GE | 2.7%EC 2.7%GE | 0.667 | 0.571 | *0.053* | **1.359** |

Table 6. The comparison of various distances used to evaluate the advantage of the two projects with respect to the baseline. Also, the DMR project is directly compared to the MR project. The proposed graph-based distance shows a promising performance in the form of its sensitivity to both 'structural' and also 'weight' differences in each comparison. In contrast, the traditional LCA aggregation of footprint fails to observe highly-sensitive but low-footprint bottlenecks, and the traditional graph distance is mainly focused on the structural differences and ignores the weights.

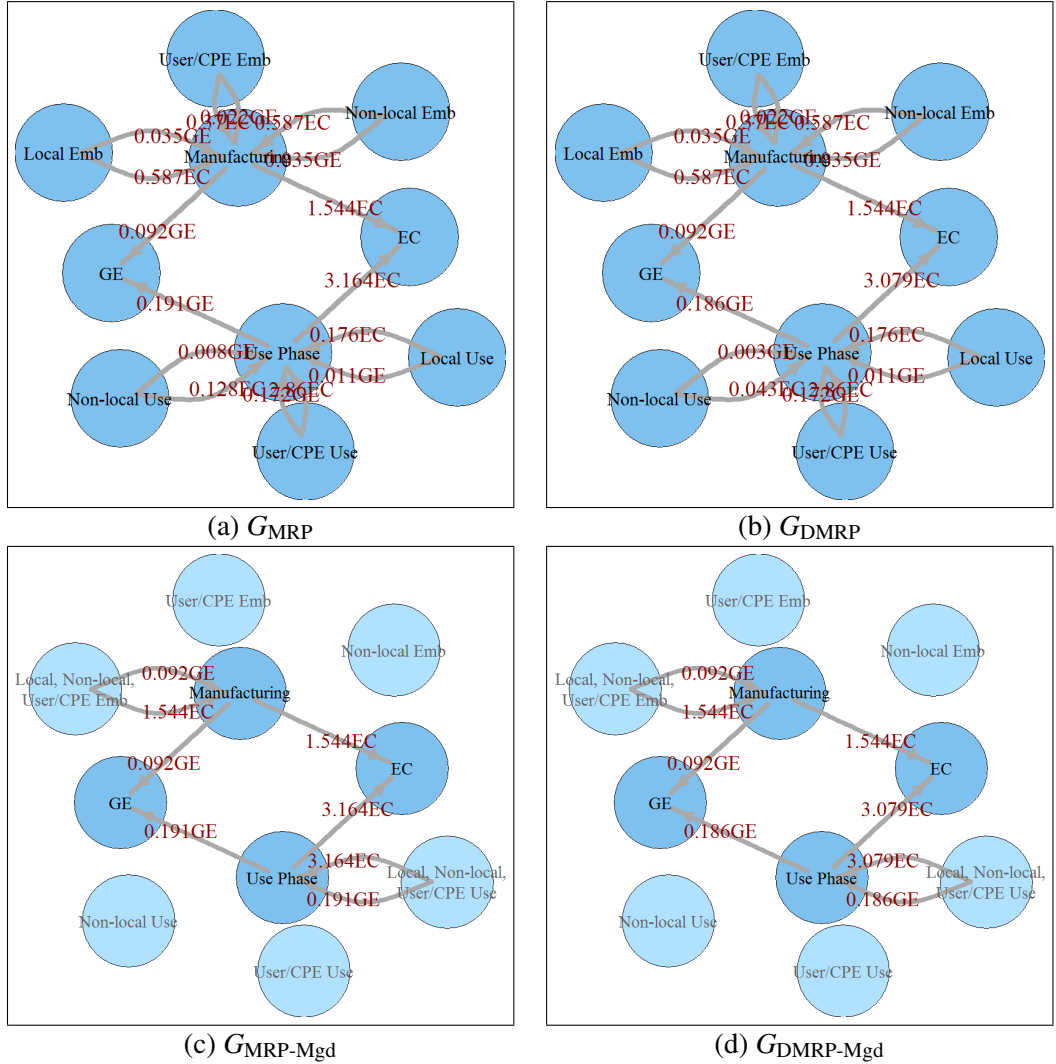

Figure 5. Compared to Figure 4, the footprint graph of the project scenarios in our case study. The left column corresponds to the Micro-registration Project, while the right column is associated with the Delayed Micro-registration Project. a)-b) The footprint graphs, and c)-d) The associated reduced graphs after applying the graph-merge operation.

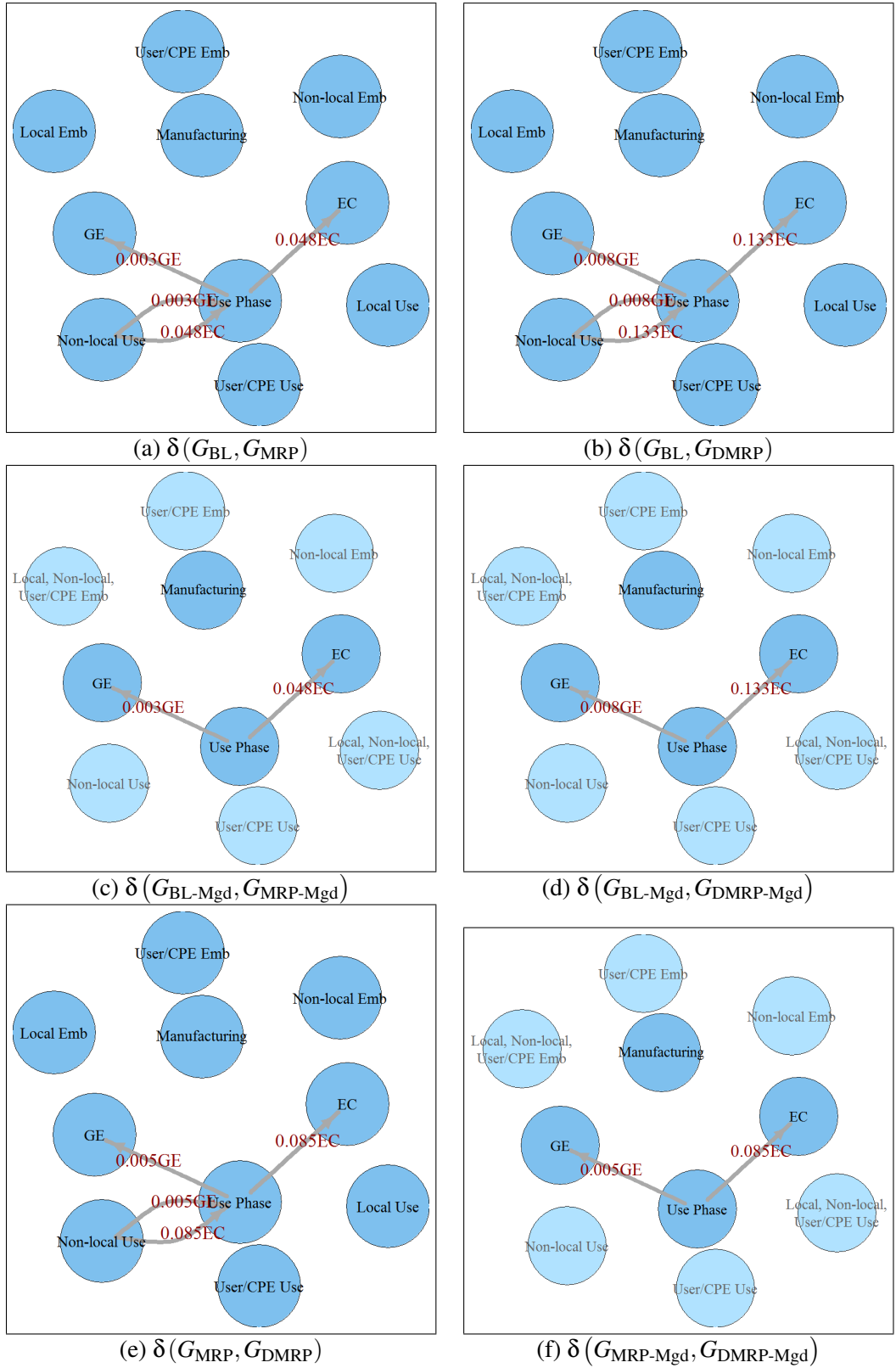

Figure 6. The difference graphs. a)-b) The differential graphs with respect to the baseline scenario, c)-d) The differential graphs with respect to the baseline scenario (for the merged graphs), e) The differential graph between the Delayed Micro-Reg and Micro-Reg projects, and f) The same as (e) but for their merged graphs.

## 9. Conclusions

In continuation of application of assessment approaches to the complex and evolving ICT cases, a graph-based distance function is proposed to compare the advantage of a project with respect to a baseline or another project. The role of small but critical parts and components is preserved by considering and respecting geographical and other attributes that can be assigned to these parts. In addition, an attribute-sensitive graph merge operation is proposed to enable reduction of the complex graphs while respecting the attributes. The proposed graph-based distance and merge functions are used in a case study related to online content services, and it is shown that these functions could simply enable the assessment to recognize the low-footprint bottleneck of the system, i.e., the backhaul connection of the local ISP in the case study considered in this work. In addition, it is shown that the two proposed projects not only postpone a high footprint regional fiber deployment, they also increase the quality of service of the delivered content by registering and synchronizing service requests.

In future, generalization of the proposed approaches to the case of IP communications, especially video calls, will be considered. In addition, the potential of 'inclusion' as a means to improve the efficiency and engagement while reducing footprint will be explored. On another front, we will generalize and use multi-layer graphs in analysis of OnCSs along location, i.e, retail ISP vs. customer premises, and also along time, for example 3-hours prime time vs. all-day time.

## Acknowledgments

The authors thank the NSERC of Canada for their financial support under Grant CRDPJ 424371-11 and also under the Canada Research Chair in Sustainable Smart Eco-Cloud. We are grateful to Ms. Farimah Farrahi Moghaddam for accepting and carrying out a very careful and detailed editing of this paper.

## A  An Example of How to Calculate the Graph-based Distances

In this appendix, we calculate the proposed graph-based distance between the baseline and the micro-registration project for both cases of full graphs and merged graphs. We start with the full graphs.

### A1. Graph-based distance between $G_{\text{BL}}$ and $G_{\text{MRP}}$

To calculate $\|\delta(G_{\text{BL}}, G_{\text{MRP}})\|_{\Omega, G_{\text{BL}}}$, we need to calculate the relative vertex weight of all vertices in Figure 6(a):

$$w^{\alpha}_{\text{vertex},G_{\text{BL}},\text{Local Use}} = 0, \quad w^{\alpha}_{\text{vertex},G_{\text{BL}},\text{User/CPE Use}} = 0, \quad w^{\alpha}_{\text{vertex},G_{\text{BL}},\text{Local Emb}} = 0,$$
$$w^{\alpha}_{\text{vertex},G_{\text{BL}},\text{Non-local Emb}} = 0, \quad w^{\alpha}_{\text{vertex},G_{\text{BL}},\text{User/CPE Emb}} = 0, \quad w^{\alpha}_{\text{vertex},G_{\text{BL}},\text{Manufacturing}} = 0,$$
$$w^{\alpha}_{\text{vertex},G_{\text{BL}},\text{EC}} = 0, \quad w^{\alpha}_{\text{vertex},G_{\text{BL}},\text{GE}} = 0,$$
$$w^{\text{EC}}_{\text{vertex},G_{\text{BL}},\text{User Phase}} = \left(\tfrac{0.048}{3.212}\right)\left(w^{\text{EC}}_{\text{vertex},G_{\text{BL}},\text{EC}} + 1\right) = \left(\tfrac{0.048}{3.212}\right)(0+1) = 0.015,$$
$$w^{\text{GE}}_{\text{vertex},G_{\text{BL}},\text{User Phase}} = \left(\tfrac{0.003}{0.194}\right)\left(w^{\text{GE}}_{\text{vertex},G_{\text{BL}},\text{GE}} + 1\right) = \left(\tfrac{0.003}{0.194}\right)(0+1) = 0.016,$$
$$w^{\text{EC}}_{\text{vertex},G_{\text{BL}},\text{Non-local Use}} = \left(\tfrac{0.048}{0.176}\right)\left(w^{\text{EC}}_{\text{vertex},G_{\text{BL}},\text{User Phase}} + 1\right) = \left(\tfrac{0.048}{0.176}\right)(0.015+1) = 0.277,$$
$$w^{\text{GE}}_{\text{vertex},G_{\text{BL}},\text{Non-local Use}} = \left(\tfrac{0.003}{0.011}\right)\left(w^{\text{GE}}_{\text{vertex},G_{\text{BL}},\text{User Phase}} + 1\right) = \left(\tfrac{0.003}{0.011}\right)(0.016+1) = 0.277.$$

Now, the vertex part of the distance can be easily calculated:

$$\|\delta(G_{\text{BL}}, G_{\text{MRP}})\|_{\text{vertex},\Omega,G_{\text{BL}}} = 0.015 + 0.016 + 0.277 + 0.277 = 0.585.$$

The edge part of the distance is more straightforward:

$$\begin{aligned}
\|\delta(G_{\text{BL}}, G_{\text{MRP}})\|_{\text{edge},\Omega,G_{\text{BL}}} &= \left(\frac{w^{\text{EC}}_{\text{Use Phase,EC}}}{w^{\text{EC}}_{G_{\text{BL}},\text{Use Phase,EC}}}\right) + \left(\frac{w^{\text{GE}}_{\text{Use Phase,GE}}}{w^{\text{GE}}_{G_{\text{BL}},\text{Use Phase,GE}}}\right) \\
&\quad + \left(\frac{w^{\text{EC}}_{\text{Non-local Use,Use Phase}}}{w^{\text{EC}}_{G_{\text{BL}},\text{Use Phase,Use Phase}}}\right) + \left(\frac{w^{\text{GE}}_{\text{Non-local Use,Use Phase}}}{w^{\text{GE}}_{G_{\text{BL}},\text{Use Phase,Use Phase}}}\right) \\
&= \left(\frac{0.048}{3.212}\right) + \left(\frac{0.003}{0.194}\right) + \left(\frac{0.048}{0.176}\right) + \left(\frac{0.003}{0.011}\right) \\
&= 0.576.
\end{aligned}$$

That means:

$$\begin{aligned}
\|\delta(G_{\text{BL}}, G_{\text{MRP}})\|_{\Omega,G_{\text{BL}}} &= \left(\|\delta(G_{\text{BL}}, G_{\text{MRP}})\|_{\text{vertex},\Omega,G_{\text{BL}}} + \|\delta(G_{\text{BL}}, G_{\text{MRP}})\|_{\text{edge},\Omega,G_{\text{BL}}}\right) \\
&= (0.585 + 0.576) = 1.161.
\end{aligned}$$

Therefore, the final graph-based distance between $G_{\text{BL}}$ and $G_{\text{MRP}}$ is:

$$\begin{aligned}
d_{\Omega}(G_{\text{BL}}, G_{\text{MRP}}) &= \tfrac{1}{2}\left(\|\delta(G_{\text{BL}}, G_{\text{MRP}})\|_{\Omega,G_{\text{BL}}} + \|\delta(G_{\text{MRP}}, G_{\text{BL}})\|_{\Omega,G_{\text{MRP}}}\right) \\
&= \tfrac{1}{2}(1.161 + 0) = 0.581.
\end{aligned} \qquad (19)$$

### A2. Graph-based distance between $G_{\text{BL-Mgd}}$ and $G_{\text{MRP-Mgd}}$

The same procedure as that of previous subsection can be followed to calculate $\|\delta(G_{\text{BL-Mgd}}, G_{\text{MRP-Mgd}})\|_{\Omega,G_{\text{BL-Mgd}}}$ but this time with the difference graph provided in Figure 6(c):

$$w^{\alpha}_{\text{vertex},G_{\text{BL-Mgd}},\text{Local Use}} = 0, \quad w^{\alpha}_{\text{vertex},G_{\text{BL-Mgd}},\text{User/CPE Use}} = 0, \quad w^{\alpha}_{\text{vertex},G_{\text{BL-Mgd}},\text{Local Emb}} = 0,$$

$$w^{\alpha}_{\text{vertex},G_{\text{BL-Mgd}},\text{Non-local Emb}} = 0, \quad w^{\alpha}_{\text{vertex},G_{\text{BL-Mgd}},\text{User/CPE Emb}} = 0, \quad w^{\alpha}_{\text{vertex},G_{\text{BL-Mgd}},\text{Manufacturing}} = 0,$$

$$w^{\alpha}_{\text{vertex},G_{\text{BL-Mgd}},\text{EC}} = 0, \quad w^{\alpha}_{\text{vertex},G_{\text{BL-Mgd}},\text{GE}} = 0, \quad w^{\alpha}_{\text{vertex},G_{\text{BL-Mgd}},\text{Non-local Use}} = 0,$$

$$w^{\text{EC}}_{\text{vertex},G_{\text{BL-Mgd}},\text{User Phase}} = \left(\tfrac{0.048}{3.212}\right)\left(w^{\text{EC}}_{\text{vertex},G_{\text{BL-Mgd}},\text{EC}}+1\right) = \left(\tfrac{0.048}{3.212}\right)(0+1) = 0.015,$$

$$w^{\text{GE}}_{\text{vertex},G_{\text{BL-Mgd}},\text{User Phase}} = \left(\tfrac{0.003}{0.194}\right)\left(w^{\text{GE}}_{\text{vertex},G_{\text{BL-Mgd}},\text{GE}}+1\right) = \left(\tfrac{0.003}{0.194}\right)(0+1) = 0.016.$$

Now, the vertex part of the distance can be easily calculated:

$$\left\|\delta\left(G_{\text{BL-Mgd}}, G_{\text{MRP-Mgd}}\right)\right\|_{\text{vertex},\mathbf{\Omega},G_{\text{BL-Mgd}}} = 0.015 + 0.016 = 0.031.$$

The edge part of the distance is more straightforward:

$$\left\|\delta\left(G_{\text{BL-Mgd}}, G_{\text{MRP-Mgd}}\right)\right\|_{\text{edge},\mathbf{\Omega},G_{\text{BL-Mgd}}} = \left(\frac{w^{\text{EC}}_{\text{Use Phase,EC}}}{w^{\text{EC}}_{G_{\text{BL-Mgd}},\text{Use Phase,EC}}}\right) + \left(\frac{w^{\text{GE}}_{\text{Use Phase,GE}}}{w^{\text{GE}}_{G_{\text{BL-Mgd}},\text{Use Phase,GE}}}\right)$$

$$= \left(\frac{0.048}{3.212}\right) + \left(\frac{0.003}{0.194}\right) = 0.030.$$

We have:

$$\left\|\delta\left(G_{\text{BL-Mgd}}, G_{\text{MRP-Mgd}}\right)\right\|_{\mathbf{\Omega},G_{\text{BL-Mgd}}} = \Big(\left\|\delta\left(G_{\text{BL-Mgd}}, G_{\text{MRP-Mgd}}\right)\right\|_{\text{vertex},\mathbf{\Omega},G_{\text{BL-Mgd}}}$$
$$+ \left\|\delta\left(G_{\text{BL-Mgd}}, G_{\text{MRP-Mgd}}\right)\right\|_{\text{edge},\mathbf{\Omega},G_{\text{BL-Mgd}}}\Big)$$
$$= (0.031 + 0.030) = 0.061.$$

Therefore, the final graph-based distance between $G_{\text{BL-Mgd}}$ and $G_{\text{MRP-Mgd}}$ is:

$$\begin{aligned} d_{\mathbf{\Omega}}\left(G_{\text{BL-Mgd}}, G_{\text{MRP-Mgd}}\right) &= \tfrac{1}{2}\Big(\left\|\delta\left(G_{\text{BL-Mgd}}, G_{\text{MRP-Mgd}}\right)\right\|_{\mathbf{\Omega},G_{\text{BL-Mgd}}} \\ &\quad + \left\|\delta\left(G_{\text{MRP-Mgd}}, G_{\text{BL-Mgd}}\right)\right\|_{\mathbf{\Omega},G_{\text{MRP-Mgd}}}\Big) \\ &= \tfrac{1}{2}(0.061 + 0) = 0.031. \end{aligned} \quad (20)$$

∎